\documentclass[%
 reprint,
superscriptaddress,
showpacs,
showkeys
amsmath,amssymb,
aps,
prl
]{revtex4-2}

\usepackage{graphicx}
\usepackage{dcolumn}
\usepackage{bm}
\usepackage{physics}
\usepackage{lipsum}
\usepackage{tikz}
\usepackage{subfigure}
\usepackage{float}
\usepackage{appendix}
\usepackage[breaklinks]{hyperref}
\usepackage{comment}
\usepackage{extarrows}
\hypersetup{colorlinks=true, linkcolor=blue, citecolor=blue, filecolor=blue, urlcolor=blue}

\begin{document}

\preprint{APS/123-QED}

\title{Hydrodynamic modes and operator spreading in a long-range\\ center-of-mass-conserving Brownian SYK model}

\author{Bai-Lin Cheng}
\affiliation{School of Physics, Peking University, Beijing 100871, China}

\author{Shao-Kai Jian}%
\email{sjian@tulane.edu}
\affiliation{Department of Physics and Engineering Physics, Tulane University, New Orleans, Louisiana 70118, USA}
\author{Zhi-Cheng Yang}
\email{zcyang19@pku.edu.cn}
\affiliation{School of Physics, Peking University, Beijing 100871, China}
\affiliation{Center for High Energy Physics, Peking University, Beijing 100871, China}

\date{\today}

\begin{abstract}
We study a center-of-mass-conserving Brownian complex Sachdev-Ye-Kitaev model with long-range (power-law) interactions characterized by $1/r^\eta$. The kinetic constraint and long-range interactions conspire to yield rich hydrodynamics associated with the conserved charge, which we reveal by computing the Schwinger-Keldysh effective action. Our result shows that charge transport in this system can be subdiffusive, diffusive, or superdiffusive, with the dynamical exponent controlled by $\eta$. 
We further employ a doubled Hilbert space methodology to derive an effective action for the out-of-time-order correlator (OTOC), from which we obtain the phase diagram delineating regimes where the lightcone is linear or logarithmic. 
Our results provide a concrete example of a quantum many-body system with kinetic constraint and long-range interactions in which the emergent hydrodynamic modes and OTOC can be computed analytically.

\end{abstract}

\maketitle

{\it Introduction.-}Understanding chaos, thermalization and emergent hydrodynamics in quantum many-body systems is one of the most significant challenges in modern quantum physics. Equilibration of conserved quantities, such as energy and particle number, is often described by a coarse-grained hydrodynamic theory that emerges at long timescales, with microscopic details of the underlying dynamics packaged into a few transport coefficients. Generically, one expects that charge relaxes diffusively at high temperatures. However, exploring what other types or universality classes of hydrodynamics can possibly emerge in quantum many-body systems has been of great interests over the past few years. For example, it has been demonstrated that imposing constraints on the allowed dynamical moves can slow down or completely freeze thermalization, leading to anomalously slow subdiffusive charge relaxation~\cite{PhysRevLett.127.230602, PhysRevResearch.2.033124, PhysRevLett.125.245303, PhysRevX.10.011042, PhysRevLett.129.150603, PhysRevE.103.022142, PhysRevB.100.214301, PhysRevB.101.214205, PhysRevB.108.144308}. On the other hand, quantum platforms such as ultracold atoms and trapped ions~\cite{RevModPhys.93.025001} naturally realize long-range interactions that decay as a power law $1/r^{\eta}$ of the separation.  
Such long-range interactions can render hydrodynamics superdiffusive, as was confirmed in experiments~\cite{doi:10.1126/science.abk2400}. The effect of the interplay between kinetic constraints and long-range interactions on transport has only been touched upon recently~\cite{morningstar2023hydrodynamics, ogunnaike2023unifying}. 
Nonetheless, explicitly computing the hydrodynamic modes in a genuine quantum many-body systems where nontrivial kinetic constraints and long-range interactions coexist has remained a nontrivial task.

Besides anomalous hydrodynamics, the dynamics of information propagation is also drastically altered in long-range interacting systems. In contrast to systems with local interactions where information can spread at most linearly~\cite{lieb1972finite}, power-law interacting systems can exhibit a much richer structure of emergent lightcone, depending on $\eta$ and the spatial dimension. 
The contour of the lightcone can be diagnosed by either generalizations of the Lieb-Robinson bound~\cite{PhysRevLett.113.030602, PhysRevLett.114.157201, else2020improved, PhysRevLett.127.160401, PhysRevX.9.031006}, or the out-of-time-order correlator (OTOC)~\cite{nahum2018operator,keyserlingk2018operator,khemani2018operator,rakovszky2018diffusive,khemani2018velocity,xu2019locality}, a key tool for studying the spreading of local operators in quantum systems. 
Recently, the phase diagram of the emergent lightcone in power-law interacting systems has been proposed based on numerical simulations and mapping to a L\'evy flight in a classical stochastic model~\cite{PhysRevB.100.064305, PhysRevLett.124.180601, PhysRevB.107.014201}. However, an understanding of operator spreading when both nontrivial kinetic constraints and long-range interactions come into play has remained limited. How will the lightcone structure be modified due to such unconventional interactions? 
A clear answer to such kind of questions is often hindered by the lack of an exactly solvable model. Therefore, it is crucial to construct solvable models and develop versatile methodologies that allow for a simultaneous study of charge transport and operator spreading, thereby providing valuable insights into the interplay between different types of interactions and their effects on thermalization and information dynamics.

In this work, we construct an analytically tractable model, namely, a Brownian complex Sachdev-Ye-Kitaev (SYK) model~\cite{sachdev1993gapless,kitaev2015simple,maldacena2016remarks,polchinski2016the,saad2018semiclassical,sunderhauf2019quantum,jian2020note,chen2020many,xu2020emergent,zhang2022quantum,zhang2023information} with power-law interactions~\cite{zhang2021universal,sahu2022entanglement} and center-of-mass conservation. The emergent hydrodynamic modes associated with particle transport can be explicitly extracted by computing the Schwinger-Keldysh effective action in the large-$N$ limit. Our result shows that charge transport in this system can be subdiffusive, diffusive, or superdiffusive, with the dynamical exponent controlled by $\eta$. To probe the contour of the emergent lightcone in this model, we further employ a doubled Hilbert space methodology to derive an effective action on a double Keldysh contour with four branches for 
the OTOC. From the effective action, we obtain the phase diagram marking regimes where the lightcone is linear and logarithmic, respectively. The phase boundary coincides with those previously proposed for generic chaotic power-law interacting systems, indicating that operator spreading is unaffected by this particular type of kinetic constraint that we impose. 

\begin{figure}
    \centering
    \includegraphics[width=1.05\linewidth]{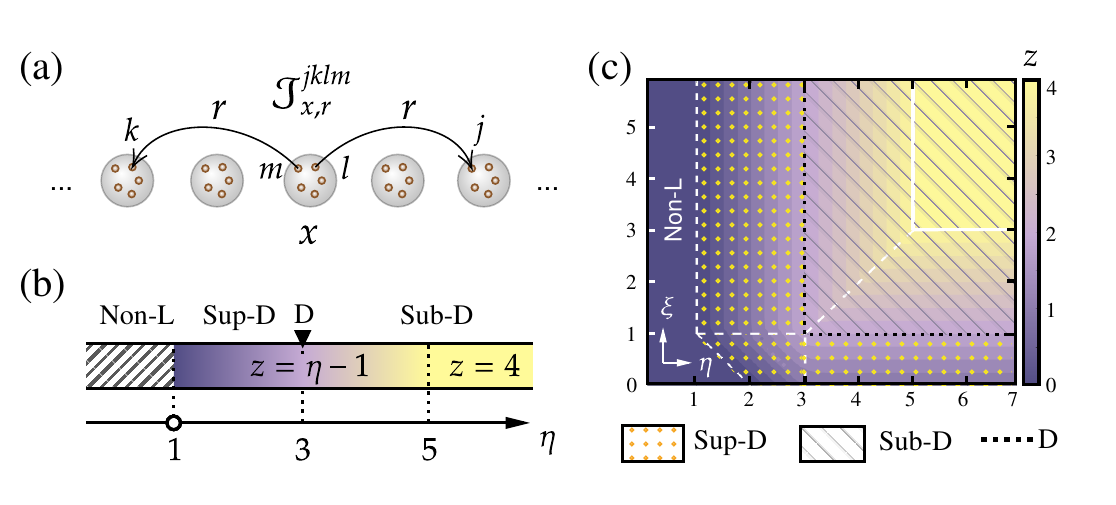}
    \caption{(a) Illustration of the complex SYK chain with center-of-mass conservation described by Hamiltonian~(\ref{eq:model}). The random couplings $\mathcal{J}^{jklm}_{x,r}$ have zero mean and their variance decays as $r^{-\eta}$. (b) The dynamical phase diagram, characterized by the hydrodynamic modes and transport behavior as a function of $\eta$. (Non-L: highly non-local dynamics, Sup-D: superdiffusive dynamics, D: diffusive dynamics, Sub-D: subdiffusive dynamics.) (c) Phase diagram of the generalized pair hopping model~(\ref{eq:model2}). The color bars used in (b)\&(c) are the same.
    }
    \label{fig1}
\end{figure}

{\it Model.-}
We consider a one-dimensional Brownian SYK chain with $N$ flavors of complex fermions on each site. The Hamiltonian consists of four-body ``pair-hopping" terms that conserve the total center-of-mass of the particles [Fig.~\ref{fig1}(a)]:
\begin{equation}
    H=\sum_{x,r}\mathcal{J}_{x,r}^{jklm}(t)\ \hat{c}^{\dagger}_{x+r,j}\hat{c}^{\dagger}_{x-r,k} \hat{c}_{x,l} \hat{c}_{x,m}+\mathrm{h.c.}\, ,
\label{eq:model}
\end{equation}
where $x$ labels the site, and $j, k, l, m =1,2,\ldots, N$ are flavor indices summed according to the Einstein summation convention. The fermion creation and annihilation operators satisfy the standard algebra $\{\hat{c}_{x,i},\hat{c}^{\dagger}_{y,j}\}=\delta_{ij}\delta_{xy}$. 
Hamiltonian~(\ref{eq:model}) describes the process where a pair of fermions on site $x$ are annihilated, followed by the creation of  another pair on sites $x\pm r$ (and the reverse process), in such a way that the total center-of-mass is preserved. 
The amplitude for such a process is encoded in the time-dependent random couplings $\mathcal{J}^{jklm}_{x,r}(t)$, with zero mean and variance
\begin{equation}
    \overline{\mathcal{J}^{jklm}_{x,r}(t_1){\mathcal{J}^{jklm}_{x,r}}^*(t_2)}=\frac{3!}{N^3}\frac{J}{r^{\eta}}\delta(t_1-t_2)\,.
\end{equation}
where $\eta$ ($J$) is the exponent (strength) of long-range couplings. The time dependence of the Hamiltonian~(\ref{eq:model}) results in the breakdown of energy conservation, implying that only the infinite temperature ensemble is well-defined. In what follows, we will focus on the large-$N$ limit of Hamiltonian~(\ref{eq:model}) so that the model becomes analytically tractable.

Since the problem is intrinsically non-equilibrium, we start by constructing the partition function living on the Keldysh contour: $Z = {\rm tr}[U \rho_0 U^\dagger]$, where the initial state $\rho_0$ is chosen to be the infinite temperature state, and $U$ is the time-evolution operator~\cite{keldysh1965diagram}. To obtain the effective action averaged over the random couplings $I= -\overline{{\rm log} Z}$, we follow the standard replica trick and integrate out the Gaussian random couplings $\{\mathcal{J}^{jklm}_{x,r}(t) \}$. Notice that since the temporal correlations of $\mathcal{J}$'s are $\delta$-functions, the resulting effective action is local on the Keldysh contour. To further decouple the eight-fermion interactions, we introduce bilocal fields: $G_x^{ab}(t_1,t_2)=\frac{1}{N}\sum_{j=1}^{N}\hat{c}^a_{x,j}(t_1)\hat{c}^{b\dagger}_{x,j}(t_2)$, where $a,b\in\{1,2\}$ label the forward $(+)$ and backward $(-)$ Keldysh contours. Integrating out the fermionic fields leads to the following effective action (see Supplemental Material (SM)~\cite{SM}):
\begin{eqnarray}
    -\frac{I}{N}
    &&=\sum_{ab,x}\mathrm{Tr}\log\qty[(-1)^{a\!+1}\delta^{ab}\partial_t-\Sigma_x^{ab}]\nonumber\\
    &&-\sum_{ab,x}\!\int\!\Sigma_x^{ba}(t_2,\!t_1)G^{ab}_x(t_1,\!t_2)\nonumber
    +\!\!\sum_{ab,xr}(-1)^{a\!+b\!+1}\frac{J}{4r^{\eta}}\\
    &&\times\!\!\int \!\!G^{ba}_{\!x\!+r}(t_2,\!t_1)G^{ba}_{\!x\!-r}(t_2,\!t_1)\qty(G_x^{ab}(t_1,\!t_2))^2\delta(t_1\!-\!t_2),\nonumber\\
    \label{eq:action}
\end{eqnarray}
where $\int\equiv \iint {\rm d}t_1 {\rm d}t_2$. The self-energy $\Sigma_x^{ab}(t_1, t_2)$ is introduced as a Lagrange multiplier to enforce the definition of the bilocal field $G^{ab}_x$. The effective action is manifestly proportional to $N$, so in the large-$N$ limit it is dominated by the saddle point plus fluctuations around the saddle point.

{\it Hydrodynamic modes.-}The saddle point equations of the action~(\ref{eq:action}) can be straightforwardly obtained by applying the variational method with respect to $G$ and $\Sigma$. The spatially uniform saddle point solution reads~\cite{SM}:
\begin{equation}\label{eq:Gt}
    \overline{G}(t)=\frac{1}{2}e^{-\frac{\Gamma}{2}|t|}\left(
    \begin{array}{cc}
        -\mathrm{sgn}(t) & 1 \\
        -1 & \mathrm{sgn}(t)
    \end{array}
    \right),
\end{equation}
where $\Gamma=\frac{J}{4}\sum_r r^{-\eta}$.
To capture the emergent hydrodynamic modes, we now consider fluctuations around the saddle point solution.
We start by writing $G$ and $\Sigma$ in terms of their saddle point values plus fluctuations: $G_x(t_1,t_2)=\overline{G}(t_1\!-\!t_2)+\delta G_x(t_1,t_2)$ and $\Sigma_x(t_1,t_2)=\overline{\Sigma}(t_1\!-\!t_2)+\delta\Sigma_x(t_1)\delta(t_1\!-\!t_2)$. Now we expand the effective action~(\ref{eq:action}) to second order in the fluctuations. The zeroth order is simply the classical value, and the first order vanishes as required by the saddle point condition. The effective action encoding Gaussian fluctuations hence takes the form~\cite{SM}: 
\begin{widetext}
\begin{eqnarray}
    -\frac{\delta I}{N}
    &&=-\frac{1}{2}\int\frac{{\rm d}\omega {\rm d}\Omega}{(2\pi)^2}\sum_{abcd} \delta\Sigma_x^{ab}(\Omega)\overline{G}^{bc}(\omega)\overline{G}^{da}(\omega+\Omega)\delta\Sigma^{cd}_x(-\Omega) -\sum_{ab}\int \frac{{\rm d}\Omega}{2\pi}\delta \Sigma_x^{ba}(\Omega)\delta G_x^{ab}(-\Omega) \nonumber \\
    &&+\!\!\sum_{x,r;a,b}(-1)^{a\!+b\!+1}\frac{J}{4r^{\eta}}\!\int \frac{{\rm d}\Omega}{2\pi} \qty[\qty(\delta G_{x+r}^{ba}\delta G_{x-r}^{ba}+\delta G_{x}^{ba}\delta G_{x}^{ba})\overline{G}^{ab}\overline{G}^{ab}\!\!+
    2\qty(\delta G_{x+r}^{ba}\delta G_{x}^{ab}+\delta G_{x-r}^{ba}\delta G_{x}^{ab})\overline{G}^{ba}\overline{G}^{ab}],
    \label{eq:fluctuations}
\end{eqnarray}
\end{widetext}
where in the second line we have suppressed the frequency dependence $\delta G(\Omega) \delta G(-\Omega)$.

To obtain the gapless hydrodynamic modes at long wavelengths, we proceed by systematically integrating out various gapped components of the fluctuating fields $\delta \Sigma^{ab}$ and $\delta G^{ab}$ in appropriate bases. The detailed calculation is a bit lengthy, and is left in the SM~\cite{SM}. In the end, we obtain the following effective action with only one component of the field $\delta G^{21}$:
\begin{equation}
    -\frac{\delta I}{N}=\sum_k\int_{\Omega}\qty(\Delta_k+\frac{\Omega^2}{\Delta_k})\qty|\delta G_{k}^{21}(\Omega)|^2,
    \label{eq:hydro_mode}
\end{equation}
where $\int_{\Omega}=\int \frac{d\Omega}{2\pi}$, $\Delta_k=(J/8)[\mathcal{R}_{\eta}(2k)-4\mathcal{R}_{\eta}(k)+3\mathcal{R}_{\eta}(0)]$ with $\mathcal{R}_{\eta}(k)=\zeta^{-1}(\eta)\sum_r\cos{kr}/r^{\eta}$, and $\zeta(z)$ is the Riemann zeta-function. The long-wavelength hydrodynamic mode $\Omega \sim k^z$ can now be directly read off from this effective action, by expanding $\Delta_k$ in powers of $k$, keeping the lowest order terms:
\begin{eqnarray}\label{eq:Deltak_p1}
    \Delta_k \approx \frac{J}{4}\qty[\frac{k^4}{4}\zeta(\eta-4)+\frac{2^{\eta}-8}{4}\sin\frac{\pi\eta}{2}\Gamma(1-\eta)|k|^{\eta-1}],
\end{eqnarray}
where $\Gamma(z)$ is the Gamma function~\cite{pole}. For $\eta >5 $, the hydrodynamic mode is dominated by $\Omega \sim k^4$ at small $k$, and hence we recover subdiffusive transport with $z=4$ for local center-of-mass conserving dynamics. For $\eta <5 $, the long-wavelength behavior of the hydrodynamic mode is instead given by $\Omega \sim k^{\eta-1}$, with a continuously varying dynamical exponent $z=\eta-1$ fully controlled by $\eta$. At $\eta = 5$, a more careful calculation yields $\Omega \sim k^4 {\rm log}k$ with logarithmic correction.
At $\eta=3$, transport behavior transitions from subdiffusive to diffusive, and then superdiffusive upon further decreasing $\eta$. The phase diagram is depicted in Fig.~\ref{fig1}(b), and is consistent with results obtained previously in Ref.~\cite{morningstar2023hydrodynamics} based on a heuristic classical hydrodynamic equation. However, we are now able to extract the hydrodynamic mode directly from the effective action. It is also evident from the above discussion how in general the spatial dimension $d$ enters. In $d$ dimensions, the line integral over $r$ is replaced by a $d$-dimensional integral, which eventually leads to $z=\eta-d$.

As a sanity check of our framework, we further compute the effective action for a Brownian {\it Majorana} SYK chain, where the complex fermionic operators in Eq.~(\ref{eq:model}) are replaced by real Majorana fields $\psi$ satisfying $\{\psi_{x,i},\psi_{y,j}\}=\delta_{ij}\delta_{xy}$. Despite the similarity of the two Hamiltonians, the physics are drastically different, since neither particle number nor energy is conserved in the Majorana chain, and one thus expects that there is no gapless hydrodynamic mode at long wavelengths. Indeed, by carrying out the same procedure of successively integrating out gapped degrees of freedom, we find that the resulting effective action has the same form as Eq.~(\ref{eq:hydro_mode}), but with a sign flip in the expression of $\Delta_k$~\cite{SM}. This implies that fluctuations of $\delta G^{21}$ are now gapped, consistent with the absence of gapless hydrodynamic mode in the Majorana case.

Hamiltonian~(\ref{eq:model}) can be further generalized to include processes where a pair of particles on sites $x$ and $x+q$ hops to sites $x-r$ and $x+q+r$ with $q\geq 0$, which also preserves the total center-of-mass~\cite{morningstar2023hydrodynamics}. The Hamiltonian in this case is given by
\begin{eqnarray}
    \label{eq:model2}
    &&H=\sum_{x,rq}\mathcal{J}_{x,r,q}^{jklm}(t) \hat{c}^{\dagger}_{x\!+q\!+r,j} \hat{c}^{\dagger}_{x\!-r,k}\hat{c}_{x\!+q,l} \hat{c}_{x,m}+\mathrm{h.c.},\\
    && \overline{\mathcal{J}^{jklm}_{x,r,q}(t_1){\mathcal{J}^{jklm}_{x,r,q}}^*(t_2)}=\frac{3!}{N^3}\frac{J}{r^{\eta}(q+1)^{\xi}}\delta(t_1-t_2).
\end{eqnarray}
A similar calculation again yields the effective action of the form~(\ref{eq:hydro_mode}), with $\Delta_k\!\approx \!\beta_1k^{\eta\!+\xi\!-2}\!+\beta_2k^{\eta\!-1}\!+\beta_3k^{\xi\!+1}\!+\beta_4k^4$, where the prefactors $\beta_i$ are functions of $\eta$ and $\xi$~\cite{SM}. The phase diagram on the $\eta$-$\xi$ plane is depicted in Fig.~\ref{fig1}(c), which is again consistent with the one obtained in Ref.~\cite{morningstar2023hydrodynamics} based on classical hydrodynamic equations.

\begin{figure}[tbh]
\centering
    \includegraphics[width=1.03\linewidth]{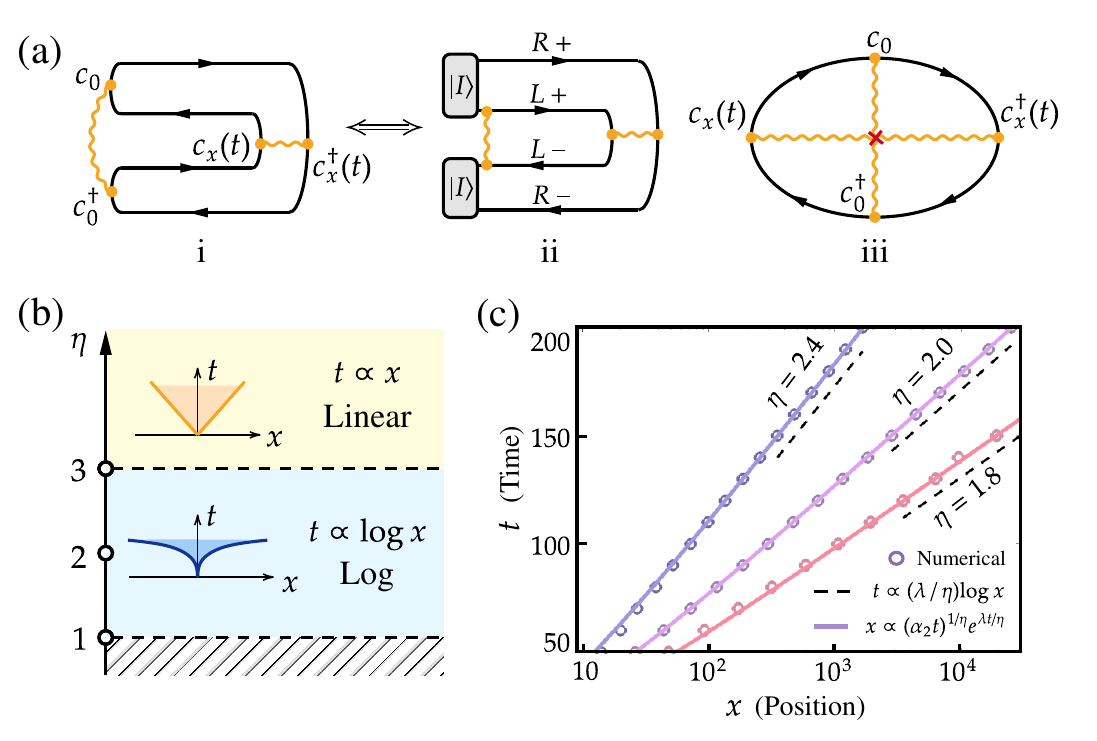}
\caption{(a) i. The OTOC~(\ref{eq:otoc}) involves four Keldysh contours (two forward and two backward); ii. This can be alternatively viewed as two Keldysh contours of a doubled Hilbert space, which are connected via a maximally entangled initial state at $t=0$; iii. A topologically equivalent diagram illustrating that the correlation function is indeed out-of-time ordered.
(b) The phase diagram of the late-time light-cone structure.
(c) Numerical calculation (hollow circle) of light-cone structures for different $\eta$ from Eq.~(\ref{eq:corr-anti}) with $J=0.1$. The analytical solution $x\!\propto\! (\alpha_2 t)^{1\!/\eta}e^{\lambda t\!/\eta}$ and the late-time asymptotic $t\!\propto\!(\lambda/\eta)\!\log x$ are plotted as solid and dashed lines, respectively.}
\label{fig2}
\end{figure}

{\it Effective action for OTOC.-} Next we compute the OTOC for Hamiltonian~(\ref{eq:model}) averaged over fermion flavors:
\begin{equation}
C(x,t) = \frac{1}{N^2}\sum_{j,k}{\rm Re}\left({\rm Tr}[\hat{c}_{x,j}(t) \hat{c}_{0,k}^\dagger \hat{c}_{x,j}^\dagger(t) \hat{c}_{0,k}] \right).
\label{eq:otoc}
\end{equation}
This quantity arises when evaluating the squared norm of the fermionic anticommutators: $\frac{1}{N^2} \sum_{j,k} {\rm Tr}\left( \{\hat{c}_{0,k}, \hat{c}_{x,j}^\dagger(t)\}^\dagger \{\hat{c}_{0,k}, \hat{c}_{x,j}^\dagger(t)\} \right)$, and thus captures the light-cone of information propagation under the evolution. Our strategy for computing the OTOC is to represent $C(x,t)$ as a two-point correlation function of the bilocal field $G$ in a doubled Hilbert space, which can then be evaluated by deriving an effective action for $G$ on four Keldysh contours (two for each copy of the Hilbert space), as illustrated in Fig.~\ref{fig2}(a) i.

We define a doubled Hilbert space by considering an identical copy of the original SYK chain, and constructing a maximally entangled state between the two copies (which we label as $L$ and $R$) on each site~\cite{choi1975completely,jamiolkowski1972linear,gu2017spread, qi2019quantum}:
\begin{equation}
|I\rangle = \bigotimes_{x} \frac{|10\rangle_{LR} - i |01\rangle_{LR}}{\sqrt{2}},
\label{eq:state}
\end{equation}
where from now on we suppress the flavor indices to lighten the notation. State~(\ref{eq:state}) has the following properties that will turn out to be useful for our purpose: (i) tracing out system $R$ leads to an infinite temperature density matrix for system $L$; (ii) $\hat{c}_L|I\rangle = i \hat{c}_R|I\rangle$, $\hat{c}_L^\dagger |I\rangle = i \hat{c}_R^\dagger|I\rangle$; 
(iii) $H_{\!L}|I\rangle = H_{\!R} |I\rangle$. Property (ii) can be viewed as a boundary condition for the fermionic operators acting on the two copies given the initial state $|I\rangle$. The OTOC~(\ref{eq:otoc}) involves two forward and two backward contours; alternatively, this can be viewed as a {\it single} forward and backward contour of a {\it doubled} Hilbert space, starting from the initial state $|I\rangle$, as depicted in Fig.~\ref{fig2}(a) ii. 
More precisely, we show in SM that OTOC of the form ${\rm Tr}[\hat{c}(t) \hat{c}^\dagger \hat{c}^\dagger(t) \hat{c}]$ is equivalent to $\langle I|\hat{c}_L^\dagger \hat{c}_L(t) \hat{c}_R^\dagger(t) \hat{c}_L |I\rangle \propto\left\langle \delta G_x^{[L\!+, R\!+]}(t) \delta G^{[L\!+,L\!-]}_0(0)\right\rangle$ which is defined as $\left\langle I\left|\mathcal{T}\!\left\{\delta G_x^{[L\!+, R\!+]}(t) \delta G^{[L\!+,L\!-]}_0(0) \right\}\right|I\right\rangle$, where $\mathcal{T}$ denotes contour ordering, $\hat{c}_L(t) \hat{c}_R^\dagger(t)= e^{i(H_{\!L}\!-H_{\!R})t} \hat{c}_L \hat{c}_R^\dagger e^{-i(H_{\!L}\!-H_{\!R})t}$~\cite{SM}.
The precise definition of $\delta G$ will be clear later. 
We thus construct the effective action on a doubled Hilbert space following the same procedure, and obtain 
\begin{eqnarray}\label{eq:action-OTOC}
    -\frac{I}{N}&&=\sum_{x,a}\mathrm{Tr}\log\qty[(-1)^a\partial_t-\Sigma_x]-\sum_{\alpha,\beta}\int\Sigma_x^{\beta\alpha}G_x^{\alpha\beta}\nonumber\\
    &&-\!\!\sum_{x,r,\alpha\beta}\!\frac{J}{4r^{\eta}}\!\!\int \!T_{\alpha\beta}G_{x\!+r}^{\beta\alpha}G_{x\!-r}^{\beta\alpha}\qty(G_{x}^{\alpha\beta})^{\!2}\delta(t_1\!-t_2)\, ,
\end{eqnarray}
where $G$ and $\Sigma$ are now $4 \times 4$ matrices, with $\alpha, \beta \equiv (s,a)$, $a=1,2$ labels the forward and backward contours, and $s=L, R$ labels the two copies of the Hilbert space. We have also introduced a sign matrix $T_{sa,\overline{s}\overline{a}}$ which is equal to $-1$ if only one of the indices $(s,\overline{s})$ or $(a,\overline{a})$ is different, and $+1$ otherwise.

The effective action~(\ref{eq:action-OTOC}) can be similarly analyzed by considering its saddle point plus Gaussian fluctuations. The saddle point solution obtained from action~(\ref{eq:action-OTOC}) is~\cite{SM}
\begin{equation}
    \overline{G}(t)=\frac{1}{2}e^{-\!\frac{\Gamma}{2}|t|}\!\left(
\begin{array}{cccc}
    \mathrm{sgn}(t) & \!\!i & \!\!\!-1 & \!\!\!i \\
    -i & \!\!\mathrm{sgn}(t) & \!\!\!-i & \!\!\!-1\\
    1 & \!\!i & \!\!\!-\mathrm{sgn}(t) & \!\!\!i \\
    -i & \!\!1 & \!\!\!-i & \!\!\!-\mathrm{sgn}(t)\\
\end{array}
    \right).
\end{equation}
For the fluctuations, it is useful to first group the sixteen components of fields $G$ symmetric and anti-symmetric parts: $\delta{G}=\left(\delta G^{\alpha\!\alpha},\frac{\delta G^{\rho\!\sigma}+\delta G^{\sigma\!\rho}}{\sqrt{2}},\frac{\delta G^{\rho\!\sigma}-\delta G^{\sigma\!\rho}}{\sqrt{2}}\right)\equiv\qty(\delta G^{\alpha\alpha},\delta G^{\{\rho,\sigma\}},\delta G^{[\rho,\sigma]})$, and similarly for $\Sigma$. Under this basis transformation, the symmetric and anti-symmetric part of the effective action~(\ref{eq:action-OTOC}) completely decouple, and can thus be handled separately~\cite{SM}.

We study the anti-symmetric part here. One can show that the OTOC~(\ref{eq:otoc}) is related to the following correlation function in the doubled Hilbert space~\cite{SM}
\begin{equation}
C(x,t) = 2i\left\langle \delta G_x^{[L+, R+]}(t) \delta G_0^{[L+,L-]}(0) \right\rangle.
\end{equation}
Starting from the anti-symmetric part of the effective action~(\ref{eq:action-OTOC}), we again successively integrate out gapped degrees of freedom of the fluctuating fields $\delta G$, and eventually end up with the following effective action involving only $\delta G^{[L+, R+]}$ and $\delta G^{[L+,L-]}$:
\begin{equation}
    -\frac{\delta I}{N}=2\sum_k\int_{\Omega}\!\delta G^{[L\!+,L\!-]}_k(\Omega)\left(i\Omega-\widetilde{\Delta}_k\right)\delta G^{[L\!+,R\!+]}_{\!-k}(-\Omega),
\end{equation}
where $\widetilde{\Delta}_k=(J/8)[\mathcal{R}_{\eta}(2k)+4\mathcal{R}_{\eta}(k)-\mathcal{R}_{\eta}(0)]$.
From the effective action above, it is straightforward to read-off the correlator that leads to the desired OTOC:
\begin{eqnarray}\label{eq:corr-anti}
   \left[\partial_t\!-\!\frac{J}{2}\zeta(\eta)\!+\alpha_1k^2\!+\alpha_2|k|^{\eta\!-\!1}\right]C(k,t)\!=\! \delta(t),
\end{eqnarray}
where we keep the first few powers of $k$ in the long-wavelength limit with coefficients $\alpha_1\!=\!J\zeta(\eta-2)/2$ and $\alpha_2\!=\!-J\Gamma(1\!-\!\eta)(2^{\eta-4}+1/2)\sin(\pi\eta/2)$. The light-cone structures with different $\eta$ can be solved according to this equation, which are depicted in the phase diagram Fig. \ref{fig2}(b).
For $\eta> 3$, the long-wavelength behavior is controlled by the $k^2$ term, and we obtain
$C(x,t)\propto \exp\qty(\lambda t-x^2/(4\alpha_1t))$.
This solution has a linear light-cone with Lyapunov exponent $\lambda=J\zeta(\eta)/2$ and butterfly velocity $v_B=\sqrt{4\lambda\alpha_1}=J\sqrt{\zeta(\eta)\zeta(\eta-2)}$. Specifically, in the regime where $\eta\gtrsim5$, the Lyapunov exponent and butterfly velocity saturate at $\lambda=J/2$ and $v_B=J$, respectively, which are consistent with Refs.~\cite{chen2020many,zhang2023information}.

For $\eta<3$, Eq.~(\ref{eq:corr-anti}) is instead dominated by $|k|^{\eta-1}$, which means that the equation involves a space-fractional derivative. We consider an instantaneous point source at the origin and solve the equation as $C(x,t)\propto e^{\lambda t}(\alpha_2 t)^{-\frac{1}{\eta-1}}Q_{\eta}(u)$, where $u=x/(\alpha_2 t)^{\frac{1}{\eta-1}}$, and 
$Q_{\eta}(u) = \int_0^{\infty}\!d\kappa\, e^{-\kappa^{\eta\!-\!1}}\!\cos{\kappa u} $ is the Fourier transformation of $e^{-|\kappa|^{\eta\!-\!1}}$. This integral is also known as the L\'{e}vy distribution function, which has a long history of research~\cite{montroll1984levy,chung1991time,wuttke2012laplace}. A key result is that the leading order of the large-$u$ expansion of $Q_{\eta}(u)$ is $u^{-\eta}$, leading to a logarithmic light-cone $\lambda t\sim\eta\log x$ in the late-time limit. A more precise light-cone solution is  $x\sim (\alpha_2 t)^{1/\eta}e^{\lambda t/\eta}$~\cite{SM}. Numerical calculations of the light-cone structure, presented in Fig. \ref{fig2}(c), agree with this analytical analysis.

The transition point at $\eta=3$ coincides with those proposed previously based on numerical simulations and mapping to a classical L\'evy flight process~\cite{PhysRevB.100.064305, PhysRevLett.124.180601, PhysRevB.107.014201, eta}. 
It is worth noting that the due to the large-$N$ nature of the SYK model, light-cones for $\eta<3$ exhibit a logarithmic structure, satisfying the mean-field limit ($N\to\infty$), in contrast to the power-law light-cone phase observed in $N=1$ simulations. We also remark that since the operators considered in the OTOC~(\ref{eq:otoc}) do not overlap with the conserved charge, the behavior of $C(x,t)$ is thus unaffected by the center-of-mass conservation and transport of the conserved charge~\cite{khemani2018operator,rakovszky2018diffusive}.

{\it Summary and outlook.-} To conclude, we have developed a solvable Brownian SYK model featuring long-range couplings and conservation of the center of mass. 
We also introduced a versatile Keldysh path integral method to analyze charge transport and operator spreading. 
This model, featuring nontrivial kinetic constraints and non-local couplings, is shown to exhibit a rich behavior of charge transport and out-of-time-order correlation.
Additionally, it serves as a prototype demonstrating the versatility of our methodology, which can be extended to study other models. 
Some intriguing questions remain open. 
Beyond the Brownian limit, it would be interesting to investigate models with energy conservation, where non-Markovian effects are anticipated to play a crucial role. 
Furthermore, our approach could be adapted to study mixed states or open quantum systems.

{\it Acknowledgments.-} We thank Xiao Chen for helpful discussions. 
S.-K. J. thanks Brian Swingle, Michael Winer and Shenglong Xu for useful discussions on related topics. 
This work is supported in part by Grant No. 12375027 from the National Natural Science Foundation of China and a Peking University startup fund (B.-L. C. and Z.-C. Y.). 
The work of S.-K. J. is supported by a start-up grant and a COR Research Fellowship from Tulane University.

\bibliography{ref}

\begin{thebibliography}{53}
\bibitem [{pol()}]{pole}%
  \BibitemOpen
  \href@noop {} {}\bibinfo {note} {One may naively conclude that $\Delta_k$ is
  divergent for $\eta =2, 3$ due to the poles of the Gamma function. However,
  this is misleading since the prefactor also vanishes at these values. One can
  directly check that the result indeed remains finite for $\eta=2,
  3$.}\BibitemShut {Stop}%
\bibitem [{eta()}]{eta}%
  \BibitemOpen
  \href@noop {} {}\bibinfo {note} {Notice that our result differs from the
  phase diagram shown in Ref.~\cite{PhysRevLett.124.180601} by a factor of 2.
  This is because we take the {\it variance} of the couplings in the
  Hamiltonian to decay as $r^{-\eta}$.}\BibitemShut {Stop}%
\end{thebibliography}

\newpage
\onecolumngrid
\appendix

\subsection*{Supplemental Material for ``Hydrodynamic modes and operator spreading in a long-range
center-of-mass-conserving Brownian SYK mode"}

\subsection{Effective action and saddle point solution for the complex chain}
We consider the Hamiltonian described in the main text, which accounts for long-range hopping in a 1D complex chain with center-of-mass conservation,
\begin{equation}\label{eq:Ham-complex-inapp}
    H=\sum_{x,r}\mathcal{J}_{x,r}^{jklm}(t)\hat{c}^{\dagger}_{x+r,j}\hat{c}^{\dagger}_{x-r,k}\hat{c}_{x,l}\hat{c}_{x,m}+\mathrm{h.c.}
\end{equation}
where the symbols are defined in the main text. 
To proceed, we integrate out the random interactions, and then make use of the effective action. 
The bilocal fields and the self-energy are defined as follows
\begin{equation}\label{eq:def-Green's-inapp}
    G_x^{ab}(t_1,t_2)\equiv \frac{1}{N}\sum_{j=1}^N \hat{c}_{x,j}^a(t_1)\hat{c}_{x,j}^{b\dagger}(t_2),
\end{equation}
\vspace{-0.2cm}
\begin{equation}
    \delta\left(G_x^{ab}(t_1,t_2)-\frac{1}{N}\sum_{j=1}^N \hat{c}_{x,j}^a(t_1)\hat{c}_{x,j}^{b\dagger}(t_2)\right)=\int [\mathcal{D}\Sigma]\exp\!\!\left[\frac{N}{2}\Sigma_x^{ba}(t_1,t_2)\left(G_x^{ab}(t_1,t_2)-\frac{1}{N}\sum_{j=1}^N \hat{c}_{x,j}^a(t_1)\hat{c}_{x,j}^{b\dagger}(t_2)\right)\right].
\end{equation}
where we introduce the Keldysh contour indices $a,b\in\{1,2\}$. Then we can integrate out the Gaussian variable $\mathcal{J}_{x,r}^{jklm}(t)$ and the complex fermion field to obtain the effective action
\begin{eqnarray}\label{eq:effaction-inapp}
    -\frac{I}{N}&&
    =\sum_{ab,x}\mathrm{Tr}\log\qty[(-1)^{a+1}\delta^{ab}\partial_t-\Sigma_x^{ab}]
    -\sum_{ab,x}\int\Sigma_x^{ba}(t_2,t_1)G^{ab}_x(t_1,t_2)\nonumber
    \\
    &&+\sum_{ab,xr}(-1)^{a+b+1}\frac{J}{4r^{\eta}}
    \int G^{ba}_{x+r}(t_2,t_1)G^{ba}_{x-r}(t_2,t_1)\qty(G_x^{ab}(t_1,t_2))^2\delta(t_1-t_2),
\end{eqnarray}
where $\int=\iint dt_1dt_2$ (which is also denoted as $\int_t$ later). The saddle point equations are obtained by applying the variational method with respect to $G$ and $\Sigma$ fields, which are
\begin{equation}\label{eq:SDeq1}
    [G_x^{-1}]^{ab}(t_1,t_2)=(-1)^a \delta^{ab}\partial_t \delta\left(t_1-t_2\right)
    +\Sigma^{ab}_x\left(t_1, t_2\right),
\end{equation}
\begin{equation}\label{eq:SDeq2}
    \Sigma_x^{ab}=(-1)^{a+b+1}\sum_r\frac{J}{4r^{\eta}}\left[(G_{x-r}^{ab})^2G_{x-2r}^{ba}+(G_{x+r}^{ab})^2G_{x+2r}^{ba}+2G_{x+r}^{ab}G_{x-r}^{ab}G_x^{ba}\right]\delta(t_1-t_2),
\end{equation}
where we briefly denote $G^{ab}(t_1,t_2)/\Sigma^{ab}(t_1,t_2)$ as $G^{ab}/\Sigma^{ab}$ and $G^{ba}(t_2,t_1)$ as $G^{ba}$. 
For a spatially uniform saddle point, the Green's function does not depend on the position (i.e., $G_x^{ab}=\overline{G}^{ab}$), we can remove the $x$ index in Eqs.(\ref{eq:SDeq1},\ref{eq:SDeq2}). Then we choose the ansatz $\overline{G}^{-1}(\omega)=i\omega\sigma_z+\Gamma Y$, where $\sigma_z$ is the $z$-component of the Pauli matrices and $Y$ is a 2-by-2 matrix to be determined, along with the coefficient $\Gamma$. The matrix elements of $Y$ can be fixed by comparing with the free fermion solution at $\Gamma=0$ (see e.g.~\cite{altland2010condensed}).
Substituting this ansatz into the saddle point equations, the Green's function in the time domain is solved as
\begin{equation}\label{eq:Gt-inapp}
    \overline{G}(t)=\frac{1}{2}e^{-\frac{\Gamma}{2}|t|}\left(
    \begin{array}{cc}
        -\mathrm{sgn}(t) & 1 \\
        -1 & \mathrm{sgn}(t)
    \end{array}
    \right),\quad \Gamma=\frac{J}{4}\sum_{r=1}^L\frac{1}{r^{\eta}}\approx\frac{1}{4}J\zeta(\eta),
\end{equation}
where in the summation over $r$, we take $L$ to infinity, turning the summation into the Riemann $\zeta$-function.

\subsection{Fluctuation analysis of the complex chain}

To analyze the non-equilibrium transport properties of the system, we consider fluctuations around the saddle point $G_x^{ab}(t_1,t_2)=\overline{G}^{ab}(t_1-t_2)+\delta G_x^{ab}(t_1,t_2)$ and $\Sigma_x^{ab}(t_1,t_2)=\overline{\Sigma}^{ab}(t_1,t_2)+\delta \Sigma_x^{ab}(t_1)\delta(t_1-t_2)$, where $\overline{G}$ and $\overline{\Sigma}$ are the saddle point solutions. We then expand the effective action (\ref{eq:effaction-inapp}) and retain terms up to the second order. First, let us consider the trace-log term
\begin{equation}\label{eq:trace-log}
    \sum_{ab}\mathrm{Tr}\,\log\left[(-1)^{a+1}\delta^{ab}\partial_t-\Sigma^{ab}_x\right]
        \supset -\frac{1}{2}\int\frac{d\omega d\Omega}{(2\pi)^2}\sum_{abcd} \delta\Sigma_x^{ab}(\Omega)\overline{G}^{bc}(\omega)\overline{G}^{da}(\omega+\Omega)\delta\Sigma^{cd}_x(-\Omega),
\end{equation}
in which we used the Fourier transformation $\delta \Sigma_x^{ab}(\Omega) = \int dt e^{i \Omega t} \delta \Sigma_x^{ab}(t)$, and $\bar G^{ab}(\omega) = \int dt e^{i \omega t} \bar G^{ab}(t)$. 
All components of $\delta\Sigma_x^{ab}$ can be packaged into a four-dimensional vector $\delta\hat{\Sigma}_x=\qty(\delta\Sigma_x^{11},\delta\Sigma_x^{12},\delta\Sigma_x^{21},\delta\Sigma_x^{22})^T$. The summation above can then be expressed as $\delta\hat{\Sigma}_x(\Omega)\cdot M(\Omega)\cdot \delta\hat{\Sigma}_x(-\Omega)$ in Fourier space. The matrix $M$ has rank two. 
In order to separate out the zero modes,  we consider a new basis $\delta\hat{\Sigma}'_x=A^{-1}\delta\hat{\Sigma}_x$, with the transformation matrix
\begin{equation}
    A=\left(
\begin{array}{cccc}
 -1 & 0 & 1 & 0 \\
 0 & -1 & 0 & 1 \\
 0 & 1 & 0 & 1 \\
 1 & 0 & 1 & 0 \\
\end{array}
\right),
\end{equation}
such that the matrix $M$ is reduced to a two-dimensional matrix $m$ of the form
\begin{equation}
    m(\Omega)=\frac{1}{2(\Gamma^2+\Omega^2)}
    \left(
\begin{array}{cc}
 \Gamma & i\Omega\\
 -i\Omega & -\Gamma \\
\end{array}
\right).
\end{equation}
Therefore the trace-log term (\ref{eq:trace-log}) simplifies to $\int\frac{d\Omega}{2\pi}\hat{\sigma}_x(\Omega)\cdot m(\Omega)\cdot \hat{\sigma}_x(-\Omega)$, where $\hat{\sigma}_x$ is a two-dimensional vector inheriting the third and fourth components of $\delta\hat{\Sigma}'_x$. The first and second components of $\delta\hat{\Sigma}'_x$ are zero modes, which will impose constraints on $\delta G_x^{ab}$ as follows.

We next consider the $\Sigma G$ coupling term
\begin{equation}
    \begin{split}
    & -\sum_{ab}\int\Sigma_x^{ba}(t_2,t_1)G^{ab}_x(t_1,t_2) \supset -\sum_{ab}\int dt_1 dt_2 \delta \Sigma_x^{ba}(t_1) \delta(t_1 -t_2) \delta G_x^{ab}(t_1, t_2) = \sum_{ab}\int d t \delta \Sigma_x^{ba}(t)  \delta G_x^{ab}(t, t) \\
    &= -\sum_{ab}\int \frac{d\Omega}{2\pi}\delta \Sigma_x^{ba}(\Omega)\delta G_x^{ab}(-\Omega)=-\int \frac{d\Omega}{2\pi}\delta \hat{\Sigma}'_x(\Omega)\cdot\qty[A^T\delta \hat{G}_x(-\Omega)],
    \end{split}
\label{eq:zero}
\end{equation}
where $\delta G_x^{ab}(\Omega) = \int dt e^{i \Omega t} G_x^{ab}(t,t) $, and 
$\delta \hat{G}_x=\qty(\delta G_x^{11},\delta G_x^{21},\delta G_x^{12},\delta G_x^{22})^T$. 
Since there is no quadratic term involving the first and second components of $\delta\hat{\Sigma}'_x$ as we discussed above, these two modes can be directly integrated out from Eq.~(\ref{eq:zero}),
which leads to two constraints: $\delta G^{11}_x=\delta G_x^{22}$ and $\delta G^{12}_x=\delta G^{21}_x$. Thus the other two components of $A^T\delta \hat{G}_x$ are simplified by substituting these constraints, which become
\begin{equation}
    \hat{\phi}_x\equiv([A^T\delta \hat{G}_x]_3,[A^T\delta \hat{G}_x]_4)^T
    =
    \left(
        \sqrt{2}\delta G^{11}_x,\, \sqrt{2}\delta G_x^{21}
    \right)^T.
\end{equation}
The kernel term, which combines the trace-log term and $\Sigma G$ coupling term, is now
\begin{equation}
    -\frac{\delta I}{N}\supset \sum_x\int\frac{d\Omega}{2\pi}\qty[\hat{\sigma}_x(\Omega)\cdot m(\Omega)\cdot \hat{\sigma}_x(-\Omega)-\hat{\sigma}_x(\Omega)\cdot\hat{\phi}_x(-\Omega)].
\end{equation}
After integrating out $\hat{\sigma}_x$, the kernel term of the effective action becomes
\begin{equation}\label{eq:kernel-comp-inapp}
    -\frac{\delta I}{N}
    \supset
    -\sum_x\int _{\Omega}\hat{g}_x(\Omega)\cdot
    \left(
    \begin{array}{cc}
 \Gamma & i\Omega \\
 -i\Omega & -\Gamma\\
\end{array}\right)
\cdot\hat{g}_x(-\Omega),
\end{equation}
where $\hat{g}_x=(\delta G_x^{11},\delta G_x^{21})$ and $\int_{\Omega}=\int \frac{d\Omega}{2\pi}$.

Next we consider interactions. The Gaussian fluctuations of the interaction term is
\begin{eqnarray}\label{eq:ab-Ieff}
    -\frac{\delta I}{N}
    &=&
    \sum_{x,r;a,b}(-1)^{a\!+b\!+1}\frac{J}{4r^{\eta}}\int_t \qty[\qty(\delta G_{x+r}^{ba}\delta G_{x-r}^{ba}+\delta G_{x}^{ba}\delta G_{x}^{ba})\overline{G}_0^{ab}\overline{G}_0^{ab}+
    2\qty(\delta G_{x+r}^{ba}\delta G_{x}^{ab}+\delta G_{x-r}^{ba}\delta G_{x}^{ab})\overline{G}_0^{ba}\overline{G}_0^{ab}]\delta(t_1-t_2)\nonumber\\
    &=&\sum_{k,r}\frac{J}{4r^{\eta}}\int_{\Omega}\hat{g}_k(\Omega)\cdot\qty(e^{2ikr}-4 e^{ikr}+1)X\cdot \hat{g}_{-k}(-\Omega)\nonumber\\
    &=&\frac{J}{4}\zeta(\eta)\sum_k\int_{\Omega}\hat{g}_k(\Omega)\cdot\qty[\mathcal{R}_{\eta}(2k)-4\mathcal{R}_{\eta}(k)+\mathcal{R}_{\eta}(0)]X\cdot\hat{g}_{-k}(-\Omega),
\end{eqnarray}
where in the first line, $\delta G^{ab}(t_1,t_2)/\delta G^{ba}(t_2,t_1)$ is simply denoted as $\delta G^{ab}/\delta G^{ba}$, and $\overline{G}_0=\overline{G}(t=0)$. In the second line, we define $\hat{g}_k=\frac{1}{\sqrt{L}}\sum_x\hat{g}_xe^{-ikx}$, and the matrix $X=\mathrm{diag}(1,-1)/2$ is calculated using the initial equilibrium solution $\overline{G}_0$. In the third line, we define $\mathcal{R}_{\eta}(k)=\zeta^{-1}(\eta)\sum_{r=1}\cos{kr}/r^{\eta}$. 
Combining Eqs.~\eqref{eq:kernel-comp-inapp} and (\ref{eq:ab-Ieff}), and expanding $\mathcal{R}_{\eta}(k)$ for small $k$, we arrive at
\begin{equation}\label{eq:action-inapp-delta-omega}
    -\frac{\delta I}{N}=\sum_k\int_{\Omega}\hat{g}_k(\Omega)\cdot\left(
    \begin{array}{cc}
        -\Delta_k & -i\Omega \\
        i\Omega & \Delta_k
    \end{array}
    \right)\cdot \hat{g}_{-k}(-\Omega),
\end{equation}
where
\begin{equation}\label{eq:def-deltak}
    \Delta_k=\frac{J}{8}[\mathcal{R}_{\eta}(2k)-4\mathcal{R}_{\eta}(k)+3\mathcal{R}_{\eta}(0)]=\frac{J}{4}\qty[\frac{k^4}{4}\zeta(\eta-4)+\frac{2^{\eta}-8}{4}\sin\frac{\pi\eta}{2}\Gamma(1-\eta)|k|^{\eta-1}]+\mathcal{O}(k^6).
\end{equation}
Finally, we integrate out one component of $\hat{g}_{-k}$, which we take to the first one, and obtain the effective action involving only $[\hat{g}_k]_2=\delta G^{21}_k$:
\begin{equation}
    -\frac{\delta I}{N}=\sum_k\int_{\Omega}\qty(\Delta_k+\frac{\Omega^2}{\Delta_k})\qty|\delta G_{k}^{21}(\Omega)|^2.
    \label{eq:hydro}
\end{equation}
The hydrodynamic mode can be readily read off from the effective action, as we discussed in the main text.

\subsection{Absence of gapless hydrodynamic mode in the Majorana chain}
In this section, we follow the same analysis to study a Brownian Majorana chain. We will show that the effective action does not contain any gapless hydrodynamic mode, which is consistent with the absence of any conserved charge in this case. 

The Hamiltonian involving Majorana fermions is given by
\begin{equation}
    H=\sum_{x,r}\mathcal{J}_{x,r}^{jklm}(t)\psi_{x+r,j}\psi_{x-r,k}\psi_{x,l}\psi_{x,m}.
\end{equation}
With the bilocal fields $G_x^{ab}(t_1,t_2)\equiv \sum_j\psi^a_{x,j}(t_1)\psi^b_{x,j}(t_2)$ and self energy $\Sigma^{ab}_x(t_1,t_2)$, we obtain the effective action for the Majorana model
\begin{eqnarray}\label{eq:effaction-maj-inapp}
    -\frac{I}{N}&&
    =\sum_{ab,x}\log\mathrm{Pf}\qty[(-1)^{a+1}\delta^{ab}\partial_t-\Sigma_x^{ab}]
    +\frac{1}{2}\sum_{ab,x}\int\Sigma_x^{ab}(t_2,t_1)G^{ab}_x(t_1,t_2)\nonumber
    \\
    &&+\sum_{ab,xr}(-1)^{a+b+1}\frac{J}{8r^{\eta}}
    \int G^{ab}_{x+r}(t_1,t_2)G^{ab}_{x-r}(t_1,t_2)\qty(G_x^{ab}(t_1,t_2))^2\delta(t_1-t_2).
\end{eqnarray}

The saddle point solution and the fluctuation around the saddle point solution can be evaluated similarly.
Here, we skip the detailed derivation, and report the main results and the main difference from the complex case. 
The Majorana fluctuation kernel action takes the same form as (\ref{eq:kernel-comp-inapp}), which is
\begin{equation}\label{eq:maj-kernel}
    -\frac{2\delta I}{N}\supset \sum_x\int_{\Omega}\hat{g}_x(\Omega)\cdot\left(
    \begin{array}{cc}
        -\Gamma & -i\Omega \\
        i\Omega & \Gamma
    \end{array}
    \right)\cdot\hat{g}_x(-\Omega),
\end{equation}
where the two-dimensional vector $\hat{g}_x$ is defined as $\hat{g}_x=(\delta G_x^{11},\delta G_x^{21})$. 
Moreover, the constraints in the Majorana model are the same as that of the complex fermion model: namely,
the constraints on $\delta G$ are still $\delta G^{11}=\delta G^{22}$ and $\delta G^{12}=\delta G^{21}$. 

The essential difference arises in the interaction term (the last term of (\ref{eq:effaction-maj-inapp})) with the quadratic expansion in the fluctuations:
\begin{equation}
\begin{split}
    -\frac{2\delta I}{N}\supset &\sum_{x,r;a,b}(-1)^{a+b+1}\frac{J}{4r^{\eta}}\int_{\Omega}\{\delta G_{x+r}^{ab}(\Omega)\overline{G}^{ab}\overline{G}^{ab}\delta G_{x-r}^{ab}(-\Omega)+\delta G_{x}^{ab}(\Omega)\overline{G}^{ab}\overline{G}^{ab}\delta G_{x}^{ab}(-\Omega)\\
    &+2\delta G_{x+r}^{ab}(\Omega)\overline{G}^{ab}\overline{G}^{ab}\delta G_{x}^{ab}(-\Omega)
    +2\delta G_{x-r}^{ab}(\Omega)\overline{G}^{ab}\overline{G}^{ab}\delta G_{x}^{ab}(-\Omega)\}\\
    =&\sum_{x,r;a,b}(-1)^{a+b+1}\frac{J}{4r^{\eta}}\int_{\Omega}\{\delta G_{x+r}^{ba}(\Omega)\overline{G}^{ab}\overline{G}^{ab}\delta G_{x-r}^{ba}(-\Omega)+\delta G_{x}^{ab}(\Omega)\overline{G}^{ba}\overline{G}^{ba}\delta G_{x}^{ab}(-\Omega)\\
    &-2\delta G_{x+r}^{ba}(\Omega)\overline{G}^{ba}\overline{G}^{ab}\delta G_{x}^{ab}(-\Omega)
    -2\delta G_{x-r}^{ba}(\Omega)\overline{G}^{ab}\overline{G}^{ba}\delta G_{x}^{ab}(-\Omega)\},
\end{split}   
\end{equation}
where we used the anti-symmetric property of the saddle point solution $\overline{G}^{ab}=-\overline{G}^{ba}$ and the constraints of the fluctuation, i.e. $\delta G^{11}=\delta G^{22}$, $\delta G^{12} = \delta G^{21}$. 
Compared with the same term in the complex model (the first line of Eq. (\ref{eq:ab-Ieff})), the difference is the sign of the last two terms. 
Therefore, in the Majorana model, the interaction term of the action simplifies to
\begin{equation}\label{eq:maj-int}
    -\frac{2\delta I}{N}\supset \frac{J}{4}\zeta(\eta)\sum_k\int_{\Omega}\hat{g}_k(\Omega)\cdot\left[\mathcal{R}_{\eta}(2k)+4\mathcal{R}_{\eta}(k)+\mathcal{R}_{\eta}(0)\right]X\cdot \hat{g}_{-k}(-\Omega),
\end{equation}
where $X=\mathrm{diag}(1,-1)/2$, as defined previously. By comparison, the complex model differs only in the sign of the $4\mathcal{R}_{\eta}(k)$ term. This difference causes $\Delta_k$ in the previous section to become
\begin{equation}\label{eq:delta+inapp}
    {\Delta}_k^+=\frac{J}{8}[\mathcal{R}_{\eta}(2k)+4\mathcal{R}_{\eta}(k)+3\mathcal{R}_{\eta}(0)]=\frac{J}{4}\qty[4\zeta(\eta)\!+\!\frac{2^{\eta}\!+\!8}{4}\sin\frac{\pi\eta}{2}\Gamma(1\!-\!\eta)|k|^{\eta\!-\!1}]+\mathcal{O}(k^2),
\end{equation}
which contains a constant, implying that the fluctuation is gapped and that the gapless hydrodynamic mode is absent.

\subsection{Generalization of the complex Brownian chain}
Hamiltonian (\ref{eq:Ham-complex-inapp}) can be further generalized to include processes where a pair of particles on sites $x$ and $x+q$ hops to sites $x-r$ and $x+q+r$ with $q\geq 0$, which also preserves the total center-of-mass. The Hamiltonian in this case is given by
\begin{equation}
    \label{eq:model2-inapp}
    H=\sum_{x,rq}\mathcal{J}_{x,r,q}^{jklm}(t) \hat{c}^{\dagger}_{x\!+q\!+r,j} \hat{c}^{\dagger}_{x\!-r,k}\hat{c}_{x\!+q,l} \hat{c}_{x,m}+\mathrm{h.c.},\quad\mathrm{with}\;\;\;
    \overline{\mathcal{J}^{jklm}_{x,r,q}(t_1){\mathcal{J}^{jklm}_{x,r,q}}^*(t_2)}=\frac{3!}{N^3}\frac{J}{r^{\eta}(q+1)^{\xi}}\delta(t_1-t_2).
\end{equation}
After introducing the Green's function, the interaction term of the action now becomes
\begin{equation}
    -\frac{\widetilde{I}}{N}\supset \sum_{x,r,q}\sum_{a,b=1}^2\frac{(-1)^{a\!+b\!+1}J}{4r^{\eta}(q+1)^{\xi}}\int G_{x\!+r\!+q}^{ba}(t_2,t_1)G_{x\!-r}^{ba}(t_2,t_1)G_{x\!+q}^{ab}(t_1,t_2)G_x^{ab}(t_1,t_2)\delta(t_1-t_2).
\end{equation}
Following the same derivation as in the previous sections, we obtain the effective action for the fluctuations
\begin{equation}
    -\frac{\delta \widetilde{I}}{N}\supset \sum_{k,r,q}\frac{J}{4r^{\eta}(q+1)^{\xi}}\int_{\Omega}\hat{g}_{k}(\Omega)\cdot\left[
    (e^{ik(2r+q)}+e^{ikq})-2(e^{-ikr}+e^{-ik(r+q)})
    \right]X\cdot\hat{g}_{-k}(-\Omega).
\end{equation}
After integrating out gapped degrees of freedom, the effective action governing the hydrodynamic mode takes the same form as Eq.~(\ref{eq:hydro}), with a modified $\Delta_k$ that depends on both $\eta$ and $\xi$, and is given by
\begin{equation}\label{eq:deltawithdpm-inapp}
\begin{split}
    \Delta_k
    &=\frac{J}{4}\zeta(\eta)\zeta(\xi)+\frac{J}{8}\zeta(\eta)\zeta(\xi)\left\{\left[\mathcal{R}_{\eta}(2k)-2\mathcal{R}_{\eta}(k)+\mathcal{R}_{\eta}(0)\right]\mathcal{R}_{\xi}(k)\cos k-2\mathcal{R}_{\eta}(k)\mathcal{R}_{\xi}(0)\right\}\\
    &=\frac{J}{16}\qty[(2^\eta-4)\mathcal{K}(\eta)\mathcal{K}(\xi)|k|^{\eta+\xi-2}+(2^\eta-8)\zeta(\xi)\mathcal{K}(\eta)|k|^{\eta-1}-2\zeta(\eta-2)\mathcal{K}(\xi)|k|^{\xi+1}+\mathcal{M}(\eta,\xi)k^4]+\mathcal{O}(k^6),
\end{split}
\end{equation}
where $\mathcal{K}(x)=\Gamma(1-x)\sin(\pi x/2)$, and $\mathcal{M}(\eta,\xi)=[\zeta(\eta-2)+\zeta(\eta-4)]\zeta(\xi)+\zeta(\eta-2)\zeta(\xi-2)$. The dependence of the dynamical exponent $z$ on $\eta$ and $\xi$ can be inferred from Eq.~(\ref{eq:deltawithdpm-inapp}), as illustrated in Fig. 1(c) of the main text.

\subsection{Introduction to the doubled Hilbert space method and its application to OTOC calculations}

\begin{figure}[t]
\centering
    \includegraphics[width=0.7\linewidth]{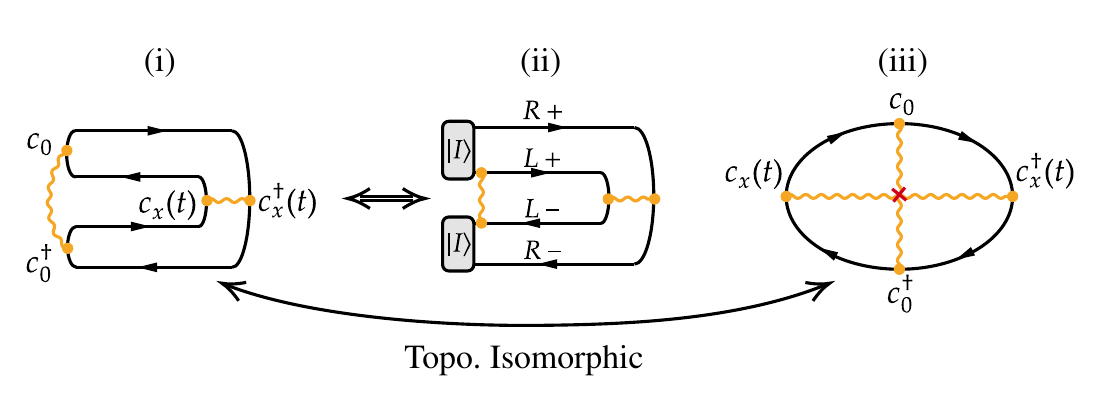}
    \caption{Keldysh contours of the doubled Hilbert space method.}
\label{fig:contour}
\end{figure}

In this section, we explain our doubled Hilbert space methodology for computing the OTOC. The OTOC as defined in the main text inevitably involves four Keldysh contours (two forward and two backward), as depicted in Fig.~\ref{fig:contour}(i). In order to derive an effective action from which the OTOC can be computed, we alternatively view the contours in Fig.~\ref{fig:contour}(i) as {\it two} Keldysh contours associated with a {\it doubled} Hilbert space (which we label as $L$ and $R$), as depicted in Fig.~\ref{fig:contour}(ii). At $t=0$, the $L$ and $R$ contours are connected via a maximally entangled EPR state $|I\rangle$ at the boundary.

As an illustration, we start by considering a Majorana chain. For simplicity, we consider Majorana operators on a single site belonging to two copies of the Hilbert space, labeled by $\psi_L$ and $\psi_R$, which can be easily generalized to the model with $2N$ flavors. The maximally entangled EPR state $|I\rangle$ at $t=0$ is most conveniently defined via the following equation~\cite{gu2017spread, qi2019quantum}:
\begin{equation}
    (\psi_L+i\psi_R)|I\rangle=0.
\end{equation}
This initial state effectively imposes a boundary condition at $t=0$ for the $L$ and $R$ operators: $\psi_L |I\rangle = -i \psi_R|I\rangle$. This also implies that $H_L|I\rangle=H_R|I\rangle$ for the Hamiltonian that involves four Majorana operators. As an exercise, let us consider the following object
\begin{equation}
\langle I | \psi_L(t) \psi_L \psi_R \psi_L(t) |I \rangle,
\end{equation}
where $\psi_L(t) = e^{iH_L t} \psi_L e^{-i H_L t}$ is the time-evolved operator in the Heisenberg picture. This quantity can in fact be represented in many equivalent ways:
\begin{eqnarray}
\langle I | \psi_L(t) \psi_L \psi_R \psi_L(t) |I \rangle &=& \langle I |e^{iH_L t} \psi_L e^{-i H_L t} \psi_L \psi_R e^{iH_L t} \psi_L e^{-i H_L t} |I\rangle \nonumber \\
&=& \langle I | \psi_L e^{-i(H_L-H_R)t} \psi_L \psi_R e^{i(H_L-H_R)t} \psi_L | I \rangle \nonumber \\
&=& \langle I |\psi_L \psi_L(t) \psi_R(t) \psi_L |I\rangle,
\end{eqnarray}
where in the second line we have used the boundary condition $H_L|I\rangle = H_R |I\rangle$. The final line can be viewed as the Heisenberg evolution under $H=H_L-H_R$ of an operator acting on both copies of the Hilbert space. Thus, in the doubled Hilbert space setup, one can similarly construct an effective action on two Keldysh contours where the time evolution is now governed by $H=H_L-H_R$.

Based on the above results,  we now show that contour-ordered correlators of the kind $\left\langle  G_x^{L+R+}(t)  G_0^{L+L-}\right\rangle$ is in fact equivalent to the OTOC originally defined on four Keldysh contours for a single copy of the Hilbert space. By definition,
\begin{eqnarray}\label{eq:OTOC-Maj-BC}
    \left\langle G_x^{L+R+}(t)G_0^{L+L-}\right\rangle&=&\left\langle I \left|{\cal{T}}\left\{\psi_{L+}(t)\psi_{R+}(t)\psi_{L+}\psi_{L-}\right\}\right|I \right\rangle\nonumber\\
    &=& \left\langle I \left|\psi_{L-}\psi_{R+}(t)\psi_{L+}(t)\psi_{L+}\right|I \right\rangle =\left\langle I \left|\psi_{L}(t)\psi_{R}\psi_{L}\psi_L(t)\right|I \right\rangle  \nonumber \\
    &\!\xlongequal{\rm BC.}& i\left\langle I \left|\psi_{L}(t)\psi_{L}\psi_{L}(t)\psi_L\right|I \right\rangle=\frac{i}{N^2}\sum_{j,k}\mathrm{Tr}\qty[\psi_{x,j}(t)\psi_{0,k}\psi_{x,j}(t)\psi_{0,k}].
\end{eqnarray}
For simplicity, we suppress the position indices and average
over fermion flavors until the last equality. The final term represents the OTO part of the squared norm of the Majorana fermionic anticommutator:
\begin{equation}
    -iC_{M}(x,t)=\frac{1}{N^2}\sum_{j,k}\mathrm{Tr}\qty(\{\psi_{0,k},\psi_{x,j}(t)\}^{\dagger}\{\psi_{0,k},\psi_{x,j}(t)\})_{\rm OTO}.
\end{equation}

Next, we apply the doubled Hilbert space method to the complex fermion model. Similar to the approach used in the Majorana model, the maximally entangled state is defined by $\hat{c}_L|I\rangle=i\hat{c}_R|I\rangle$. The form of $|I\rangle$ and the boundary condition relating $\hat{c}_L^{\dagger}$ and $\hat{c}_R^{\dagger}$ can be derived as
\begin{equation}
    \hat{c}_L|I\rangle=i\hat{c}_R|I\rangle,\quad \hat{c}_L^{\dagger}|I\rangle=i\hat{c}_R^{\dagger}|I\rangle,\quad \mathrm{with}\; |I\rangle=\bigotimes_{x} \frac{|10\rangle_{LR} - i |01\rangle_{LR}}{\sqrt{2}},
\end{equation}
where we suppress the flavor indices to lighten the notation. Using the above boundary conditions, one can also show that $H_L|I\rangle=H_R|I\rangle$. Later we will show that the OTOC for complex fermionic operators is also related to contour-ordered correlation functions on a doubled Hilbert space, similarly to Eq.~(\ref{eq:OTOC-Maj-BC}).

\subsection{OTOC calculations for the Majorana chain}
We first calculate the OTOC of the Majorana chain as a warm-up. Considering the left and right sides along with the Keldysh indices, the effective action of the Majorana chain is
\begin{equation}\label{eq_effac}
    \begin{split}
        -\frac{I}{N}=&\sum_{x}\log \mathrm{Pf} \left((-1)^a\partial_t-\Sigma_x\right)+\frac{1}{2}\sum_{\alpha,\beta}\int \Sigma^{\alpha\beta}_x(t_1,t_2)G^{\alpha\beta}_x(t_1,t_2)\\
        -&\sum_{x,r}\sum_{\alpha,\beta}\frac{J}{8r^{\eta}}\int T_{\alpha\beta}G^{\alpha\beta}_{x+r}(t_1,t_2)G^{\alpha\beta}_{x-r}(t_1,t_2)\left(G^{\alpha\beta}_x(t_1,t_2)\right)^2\delta(t_1-t_2),
    \end{split}
\end{equation}
with
\begin{equation}
    T_{\alpha\beta}=\left(
\begin{array}{cccc}
    1 & -1 & -1 & 1 \\
    -1 & 1 & 1 & -1 \\
    -1 & 1 & 1 & -1 \\
    1 & -1 & -1 & 1 \\
\end{array}
    \right),
\end{equation}
where indices $\alpha,\beta$ denote $sa$, with $s=L,R$ for the left/right side and $a=+,-$ for the Keldysh contour. We can derive the saddle point solution by considering the boundary conditions $\left(\psi_{L}+i\psi_{R}\right)|I\rangle=0$, along with the one-side solution (\ref{eq:Gt-inapp}), and obtain
\begin{equation}\label{eq:Green's-maj-otoc}
    \overline{G}(t)=\frac{1}{2}e^{-\frac{\Gamma}{2}|t|}\left(
\begin{array}{cccc}
    \mathrm{sgn}(t) & i & -1 & i \\
    -i & \mathrm{sgn}(t) & -i & -1\\
    1 & i & -\mathrm{sgn}(t) & i \\
    -i & 1 & -i & -\mathrm{sgn}(t)\\
\end{array}
    \right).
\end{equation}

We then analyze the fluctuation away from this equilibrium solution using the method similar to that in hydrodynamics. The fluctuation of the `Pf' term provides us with the kernel matrix, which decouples the diagonal and anti-diagonal parts of $\delta \Sigma^{\alpha\beta}$. We focus on the off-diagonal part, as the OTOC is only related to it. Moreover, due to the index anti-symmetry of the field, $\delta\Sigma^{\alpha\beta}=-\delta\Sigma^{\beta\alpha}$, the off-diagonal part has six components. 
Notice that the antisymmetric condition for the self energy is because the self energy is introduced to enforce the definition of bilocal field, i.e., $\prod_{ab,t_1,t_2}\delta\left(G_x^{ab} (t_1,t_2) - \frac1N \sum_j \psi_{x,j}^a(t_1) \psi_{x,j}^b(t_2) \right) = \int D\Sigma \exp\left[ \frac{N}2 \sum_{ab} \int dt_1 dt_2 \Sigma^{ab}_x(t_1,t_2) \left( G_x^{ab} (t_1,t_2) - \frac1N \sum_j \psi_{x,j}^a(t_1) \psi_{x,j}^b(t_2) \right) \right]$, and due to the antisymmetric property of the bilocal field $G_x^{ab}(t_1, t_2) = - G_x^{ba}(t_2, t_1)$, the symmetric part of $\Sigma^{ab}(t_1,t_2)$ will drop out. 
In other words, we only need to consider the antisymmetric part. 
We choose the following basis for simplicity
\begin{eqnarray}\label{eq:base-maj-inapp}
    \delta {\Sigma}^M&=&\left(\delta \Sigma^{L+L-}, \delta \Sigma^{R+R-},-i \delta \Sigma^{L+R+},-i \delta \Sigma^{L-R-},-i \delta \Sigma^{L+R-},i \delta \Sigma^{R+L-}\right),\nonumber\\
    \delta {G}^M&=&\left(\delta G^{L+L-}, \delta G^{R+R-},i \delta G^{L+R+},i \delta G^{L-R-},i \delta G^{L+R-},-i \delta G^{R+L-}\right).
\end{eqnarray}
The effective action of fluctuation can be written as
\begin{equation}\label{eq:fluc-Maj-inapp}
    -\frac{\delta I}{N}=\sum_k\int_{\Omega}\qty[\delta{\Sigma}_k(\Omega)\cdot K_M(\Omega)\cdot \delta{\Sigma}_{-k}(-\Omega)+\delta{\Sigma}_k(\Omega)\cdot\delta{G}_{-k}(-\Omega)-\delta{G}_k(\Omega)\cdot V_M(k)\cdot \delta{G}_{-k}(-\Omega)],
\end{equation}
with the diagonal interaction matrix $V_M(k)=(J/32)\qty[\mathcal{R}_{\eta}(2k)+4\mathcal{R}_{\eta}(k)+\mathcal{R}_{\eta}(0)]\mathrm{diag}(-1,-1,1,1,-1,-1)$, and the kernel matrix
\begin{equation}
    K_M(\Omega)=\frac{1}{4}\left(
    \begin{array}{cccccc}
        0 & 0 & h^* & h^* & h^* & -h^* \\
        0 & 0 & h^* & h^* & h^* & -h^* \\
        h & h & 0 & 0 & h & -h \\
        h & h & 0 & 0 & h & -h \\
        h & h & h^* & h^* & 2\Gamma|h|^2 & -2\Gamma|h|^2 \\
        -h & -h & -h^* & -h^* & -2\Gamma|h|^2 & 2\Gamma|h|^2 \\
    \end{array}
    \right),
\end{equation}
where $h=(\Gamma+i\Omega)^{-1}$ and $\Gamma=J\zeta(\eta)/4$. To reduce the rank-2 matrix $K_M(\Omega)$, we choose the basis $\hat{\sigma}=O^{-1}\delta\hat{\Sigma}$ and $\hat{g}=O^T\delta\hat{G}$ with the transformation matrix
\begin{equation}
    O=\frac{1}{2}\left(
\begin{array}{cccccc}
 -1 & 0 & 0 & 1 & 1 & 1 \\
 1 & 0 & 0 & 1 & 1 & 1 \\
 0 & 0 & \sqrt{2} & 1 & -1 & 0 \\
 0 & 0 & -\sqrt{2} & 1 & -1 & 0 \\
 0 & \sqrt{2} & 0 & -1 & -1 & 1 \\
 0 & -\sqrt{2} & 0 & -1 & -1 & 1 \\
\end{array}
\right).
\end{equation}
So that $K_M(\Omega)$ reduces to
\begin{equation}
    O^TK_MO=\left(\begin{array}{cc}
       0_{4\times4}  & 0 \\
        0 & k_M
    \end{array}\right)
    ,\quad
    k_M(\Omega)=\frac{1}{4}\left(\begin{array}{cc}
       0  & \frac{1}{\Gamma+i\Omega} \\
        \frac{1}{\Gamma-i\Omega} & -\frac{\Gamma}{\Gamma^2+\Omega^2}
    \end{array}\right).
\end{equation}
It is evident that $\{g_1,g_2,g_3,g_4\}$ are zero modes. By integrating out $\hat{\sigma}$ from Eq. (\ref{eq:fluc-Maj-inapp}), we derive the effective action, which is expressed in the original basis (\ref{eq:base-maj-inapp}) as
\begin{equation}\label{eq:OTOC-maj-eff}
    \frac{i\delta I}{N}=2\sum_k\int\frac{d\Omega}{2\pi}\delta G^{L+L-}_k(\Omega)\left(i\Omega-\widetilde{\Delta}_k\right)\delta G^{L+R+}_{-k}(-\Omega)
\end{equation}
where
\begin{equation}\label{eq:tilde-delta}
    \widetilde{\Delta}_k=\frac{J}{8}[\mathcal{R}_{\eta}(2k)+4\mathcal{R}_{\eta}(k)-\mathcal{R}_{\eta}(0)]\approx \frac{J}{2}\qty[\zeta(\eta)-\zeta(\eta-2)k^2+\Gamma(1-\eta)(2^{\eta-3}+1)\sin\frac{\pi\eta}{2}|k|^{\eta-1}].
\end{equation}
The correlator in the frequency and momentum representation, given by $C_M(\Omega,k)=\left\langle\delta G^{L+L-}_k(\Omega)\delta G^{L+R+}_{-k}(-\Omega)\right\rangle$, can be directly extracted from the action (\ref{eq:OTOC-maj-eff}), which is $C_M(\Omega,k)=\qty(i\Omega-\widetilde{\Delta}_k)^{-1}$.

\subsection{OTOC calculations for the complex chain}
The OTOC calculation for the complex chain is more tedious than in the Majorana case because the Green's function no longer exhibits index anti-symmetry, resulting in 16 components (instead of 6) that must be considered. The action of the infinite temperature complex model, with an identical copy of the original SYK chain, is given by
\begin{equation}\label{eq_effac-complex-OTOC}
    \begin{split}
        -\frac{I}{N}=&\sum_{x}\mathrm{Tr}\log  \left((-1)^a\partial_t-\Sigma_x\right)-\sum_{\alpha,\beta}\int \Sigma^{\beta\alpha}_x(t_2,t_1)G^{\alpha\beta}_x(t_1,t_2)\\
        -&\sum_{x,r}\sum_{\alpha,\beta}\frac{J}{4r^{\eta}}\int T_{\alpha\beta}G^{\beta\alpha}_{x+r}(t_2,t_1)G^{\beta\alpha}_{x-r}(t_2,t_1)\left(G^{\alpha\beta}_x(t_1,t_2)\right)^2\delta(t_1-t_2),
    \end{split}
\end{equation}
where we retain the definition of the $T_{\alpha\beta}$ matrix. As previously introduced, the boundary conditions are $\hat{c}_L|I\rangle = i \hat{c}_R|I\rangle$, $\hat{c}_L^\dagger |I\rangle = i \hat{c}_R^\dagger|I\rangle$, and $H_L|I\rangle=H_R|I\rangle$, where $|I\rangle=\bigotimes_{x} \frac{|10\rangle_{LR} - i |01\rangle_{LR}}{\sqrt{2}}$. Therefore, the saddle point solution for the Green's function is given by (\ref{eq:Green's-maj-otoc}) as well. Note that while the complex model and the Majorana model share the same saddle point solution, the fluctuation part of the former does not satisfy index anti-symmetry.
We first group the field $G$ and $\Sigma$ into symmetric (including diagonal and off-diagonal parts) and anti-symmetric parts
\begin{equation}
    \delta{G}=\left(\delta G^{\alpha\alpha},\frac{\delta G^{\rho\sigma}+\delta G^{\sigma\!\rho}}{\sqrt{2}},\frac{\delta G^{\rho\sigma}-\delta G^{\sigma\!\rho}}{\sqrt{2}}\right)\equiv\qty(\delta G^{\alpha\alpha},\delta G^{\{\rho,\sigma\}},\delta G^{[\rho,\sigma]}),
\end{equation}
\begin{equation}
    \delta{\Sigma}=\left(\delta\Sigma^{\alpha\alpha},\frac{\delta\Sigma^{\sigma\!\rho}+\delta\Sigma^{\rho\sigma}}{\sqrt{2}},\frac{\delta\Sigma^{\sigma\!\rho}-\delta\Sigma^{\rho\sigma}}{\sqrt{2}}\right)\equiv\qty(\delta\Sigma^{\alpha\alpha},\delta\Sigma^{\{\sigma,\rho\}},\delta\Sigma^{[\sigma,\rho]}),
\end{equation}
where $\alpha\in\{L-,R-,L+,R+\}$ and $\rho\sigma\in\{L+L-,R+R-,L+R+,L-R-,L+R-,R+L-\}$. Under this basis transformation, the symmetric and anti-symmetric parts of the effective action~(\ref{eq_effac-complex-OTOC}) decouple completely as shown in (\ref{eq:action-OTOC-complex-decouple}). This is indicated by the fact that the kernel matrix is partitioned into a block diagonal matrix $K=\mathrm{diag}(K_S,K_A)$, where the kernel matrix of symmetric part, $K_S$, is a 10-by-10 matrix of rank 6, and the kernel matrix of anti-symmetric part, $K_A$, is a 6-by-6 matrix of rank 2. The effective action can be written as
\begin{equation}\label{eq:action-OTOC-complex-decouple}
    -\frac{\delta I}{N}=\sum_k\int_{\Omega} \qty(\delta{\Sigma}^S)^{\!T}\! K_S\delta{\Sigma}^S-\qty(\delta{\Sigma}^S)^{\!T}\!\delta{G}^S-\qty(\delta{G}^S)^{\!T}\! V_S\delta{G}^S
    +\sum_k\int_{\Omega} \qty(\delta{\Sigma}^A)^{\!T}\! K_A\delta{\Sigma}^A-\qty(\delta{\Sigma}^A)^{\!T}\!\delta{G}^A-\qty(\delta{G}^A)^{\!T}\! V_A\delta{G}^A,
\end{equation}
where the index $S/A$ denotes the symmetric/anti-symmetric part; $\delta X^S=\qty(\delta X^{\alpha\alpha},\delta X^{\{\rho,\sigma\}})$ with $X=G,\Sigma$; $\delta G^A=\delta G^{[\rho,\sigma]}$ and $\delta \Sigma^A=\delta \Sigma^{[\sigma,\rho]}$; we suppress the $k$ and $\Omega$ dependence; $K_A=K_M$ and $V_A=V_M$, where matrices with index $M$ were introduced in the previous section. We focus only on the anti-symmetric part for reasons we will explain later.

The anti-symmetric part is described by the action with index $A$ in (\ref{eq:action-OTOC-complex-decouple}). The only difference between this action and the OTOC action for the Majorana chain described by (\ref{eq:fluc-Maj-inapp}) is the sign of the $\delta\Sigma$-$\delta G$ interaction term. However, a careful investigation of the indices shows that the minus sign is canceled since $\delta \Sigma^A=-\delta \Sigma^M$. Therefore, the OTOC action for the anti-symmetric part is similar to action (\ref{eq:OTOC-maj-eff}), except for the difference in indices: $\delta G^{\alpha\beta}$ is replaced by $\delta G^{[\alpha,\beta]}$. As a result, the correlator $C(\Omega,k)$ exhibits the same dynamics as $C_M(\Omega,k)$, which is
\begin{equation}\label{ep:corr-antisymm-def-inapp}
    C(\Omega,k)\equiv \left\langle  \delta G_{k}^{[L+,L-]}(\Omega) \delta G_{-k}^{[L+, R+]}(-\Omega) \right\rangle\propto\qty(i\Omega-\widetilde{\Delta}_k)^{-1},
\end{equation}
where $\widetilde{\Delta}_k$ is defined as (\ref{eq:tilde-delta}). Transforming the equation above to the momentum-time domain, we arrive at the equation of motion for the correlator
\begin{equation}\label{ep:corr-antisymm-inapp}
    \qty[\partial_t-\frac{J}{2}\zeta(\eta)+\alpha_1k^2+\alpha_2|k|^{\eta-1}]C(k,t)=\delta(t),
\end{equation}
where we retain the lowest powers of $k$ in the long-wavelength limit, with coefficients $\alpha_1=J\zeta(\eta-2)/2$ and $\alpha_2=-J\Gamma(1-\eta)(2^{\eta-4}+1/2)\sin(\pi\eta/2)$. One more step, Fourier transforming the correlator to the spatiotemporal domain, yields the dynamics of $C(x,t)$ and the light-cone structure.

Now, we reveal the physical interpretations of this correlator and explain why it is related to the real part of the OTOC that we are interested in. To begin with, we recall the boundary conditions
\begin{equation}
    \hat{c}_L|I\rangle = i \hat{c}_R|I\rangle,\quad \hat{c}_L^\dagger|I\rangle = i \hat{c}_R^\dagger|I\rangle,\quad H_L|I\rangle=H_R|I\rangle,
\end{equation}
and then we provide a lemma for arbitrary operators ${\cal U}$, ${\cal V}$, and $\cal O$, which is
\begin{equation}\label{eq:lemma}
    \langle I |\mathcal{U}_L^\dagger (t)\mathcal{O}_{LR}\mathcal{V}_L(t)|I\rangle=\langle I |\mathcal{U}_L^\dagger\mathcal{O}_{LR}(t)\mathcal{V}_L|I\rangle,
\end{equation}
where $\mathcal{O}_{LR}(t)=e^{i(H_L-H_R)t}\mathcal{O}_{LR}e^{-i(H_L-H_R)t}$ is the operator evolution under the doubled Hamiltonian. Indeed, since $H_L|I\rangle=H_R|I\rangle$ and $[\mathcal{A}_L,\mathcal{B}_R]=0$ for arbitrary operators $\mathcal{A}$ and $\mathcal{B}$, there is
\begin{equation}
    \mathcal{V}_L(t)|I\rangle=e^{iH_Lt}\,\mathcal{V}_L\,e^{-iH_Lt}|I\rangle=e^{i(H_L-H_R)t}\mathcal{V}_L|I\rangle.
\end{equation}


$C(x,t)$ is defined in (\ref{ep:corr-antisymm-def-inapp}) with the equation of motion (\ref{ep:corr-antisymm-inapp}). Ignoring constant and exponential decay terms, only $\left\langle G_{x}^{[L+, R+]}(t)G_{0}^{[L+,L-]}(0)  \right\rangle$ is relevant. Substituting the definition of the Green's function (\ref{eq:def-Green's-inapp}) into the correlator, we obtain
\begin{eqnarray}
    C(x,t)&\equiv& \left\langle  \delta G_{x}^{[L+, R+]}(t)\delta G_{0}^{[L+,L-]}(0)  \right\rangle\ni \left\langle  G_{x}^{[L+, R+]}(t)G_{0}^{[L+,L-]}(0)  \right\rangle\nonumber\\
    &=& \frac{1}{2}\left\langle I\left|\mathcal{T}\left\{ \left[
    \hat{c}_{L+}(t)\hat{c}_{R+}^{\dagger}(t)-\hat{c}_{R+}(t)\hat{c}_{L+}^{\dagger}(t)
    \right]\left[
    \hat{c}_{L+}\hat{c}_{L-}^{\dagger}-\hat{c}_{L-}\hat{c}_{L+}^{\dagger}
    \right]
    \right\}\right|I\right\rangle\nonumber\\
    &=&\frac{1}{2}\left\langle I\left|\hat{c}_{L-}^{\dagger}\hat{c}_{R+}^{\dagger}(t)\hat{c}_{L+}(t)\hat{c}_{L+}+\hat{c}_{L-}\hat{c}_{R+}^{\dagger}(t)\hat{c}_{L+}(t)\hat{c}_{L+}^{\dagger}+\hat{c}_{L-}^{\dagger}\hat{c}_{R+}(t)\hat{c}_{L+}^{\dagger}(t)\hat{c}_{L+}+\hat{c}_{L-}\hat{c}_{R+}(t)\hat{c}^{\dagger}_{L+}(t)\hat{c}^{\dagger}_{L+}\right|I\right\rangle\nonumber\\
    &=&\frac{1}{2}\left\langle I\left|\hat{c}_{L}^{\dagger}\hat{c}_{R}^{\dagger}(t)\hat{c}_{L}(t)\hat{c}_{L}
    +\hat{c}_{L}\hat{c}_{R}^{\dagger}(t)\hat{c}_{L}(t)\hat{c}_{L}^{\dagger}
    +\hat{c}_{L}^{\dagger}\hat{c}_{R}(t)\hat{c}_{L}^{\dagger}(t)\hat{c}_{L}
    +\hat{c}_{L}\hat{c}_{R}(t)\hat{c}^{\dagger}_{L}(t)\hat{c}^{\dagger}_{L}\right|I\right\rangle,
\end{eqnarray}
where $\mathcal{T}$ in the second line denotes contour ordering. For simplicity, we suppress the position indices and average over fermion flavors from the second line. Substituting lemma (\ref{eq:lemma}) into the derivation, we have
\begin{eqnarray}
    &=&\frac{1}{2}\left\langle I\left|\hat{c}_{L}^{\dagger}(t)\hat{c}_{R}^{\dagger}\hat{c}_{L}\hat{c}_{L}(t)
    +\hat{c}_{L}(t)\hat{c}_{R}^{\dagger}\hat{c}_{L}\hat{c}_{L}^{\dagger}(t)
    +\hat{c}_{L}^{\dagger}(t)\hat{c}_{R}\hat{c}_{L}^{\dagger}\hat{c}_{L}(t)
    +\hat{c}_{L}(t)\hat{c}_{R}\hat{c}^{\dagger}_{L}\hat{c}^{\dagger}_{L}(t)\right|I\right\rangle\nonumber\\
    &=&-\frac{i}{2}\left\langle I\left|\hat{c}_{L}^{\dagger}\hat{c}_{L}^{\dagger}(t)\hat{c}_{L}\hat{c}_{L}(t)
    +\hat{c}_{L}\hat{c}_{L}^{\dagger}(t)\hat{c}_{L}\hat{c}_{L}^{\dagger}(t)
    +\hat{c}_{L}^{\dagger}\hat{c}_{L}(t)\hat{c}_{L}^{\dagger}\hat{c}_{L}(t)
    +\hat{c}_{L}\hat{c}_{L}(t)\hat{c}^{\dagger}_{L}\hat{c}^{\dagger}_{L}(t)\right|I\right\rangle\nonumber\\
    &=& -\frac{2i}{N^2}\sum_{j,k}{\rm Re}\left({\rm Tr}[\hat{c}_{x,j}(t) \hat{c}_{0,k}^\dagger \hat{c}_{x,j}^\dagger(t) \hat{c}_{0,k}] \right),
\end{eqnarray}
where we reintroduce indices and average over fermion flavors in the last line. Finally, we have demonstrated that $C(x,t)$ indeed encodes information about the real part of the OTOC, which arises when evaluating the squared norm of the fermionic anticommutator
\begin{equation}
    \frac{i}{2}C(x,t)=\frac{1}{N^2} \sum_{j,k} {\rm Tr}\left( \{\hat{c}_{0,k}, \hat{c}_{x,j}^\dagger(t)\}^\dagger \{\hat{c}_{0,k}, \hat{c}_{x,j}^\dagger(t)\} \right)_{\rm OTO}.
\end{equation}

\subsection{Analysis and numerical solutions of the light-cone structure for general $\eta<3$}
According to the derivations in the previous section, the equation of motion for the correlator $C(k,t)$ is
\begin{equation}
    \left[\partial_t-\lambda+\alpha_1k^2+\alpha_2|k|^{\eta-1}\right]C(k,t)=\delta(t),
\end{equation}
where $\lambda=J\zeta(\eta)/2$, $\alpha_1=J\zeta(\eta-2)/2$ and $\alpha_2=-J\Gamma(1-\eta)(2^{\eta-4}+1/2)\sin(\pi\eta/2)$. We consider an instantaneous point source at the origin and solve the equation above, which is
\begin{eqnarray}
    \frac{C(x,t)}{C(0,0)}&=& e^{\lambda t}\int_{-\infty}^{+\infty} \frac{dk}{2\pi} e^{ikx}\exp\qty(-\alpha_1 t k^2-\alpha_2 t |k|^{\eta\!-\!1})
    \nonumber\\
    &=&\frac{e^{\lambda t}}{\pi(\alpha_2t)^{\frac{1}{\eta-1}}}\int_0^{\infty} \!d{\kappa}\, \exp\qty(-\alpha_1\alpha_2^{\frac{2}{1-\eta}}t^{\frac{\eta-3}{\eta-1}}\kappa^2-{\kappa}^{\eta\!-\!1})\!\cos{{\kappa}u},\quad u=\frac{x}{(\alpha_2 t)^{\frac{1}{\eta\!-\!1}}}.
\end{eqnarray}
\begin{figure}[b]
    \centering
    \includegraphics[width=0.8\linewidth]{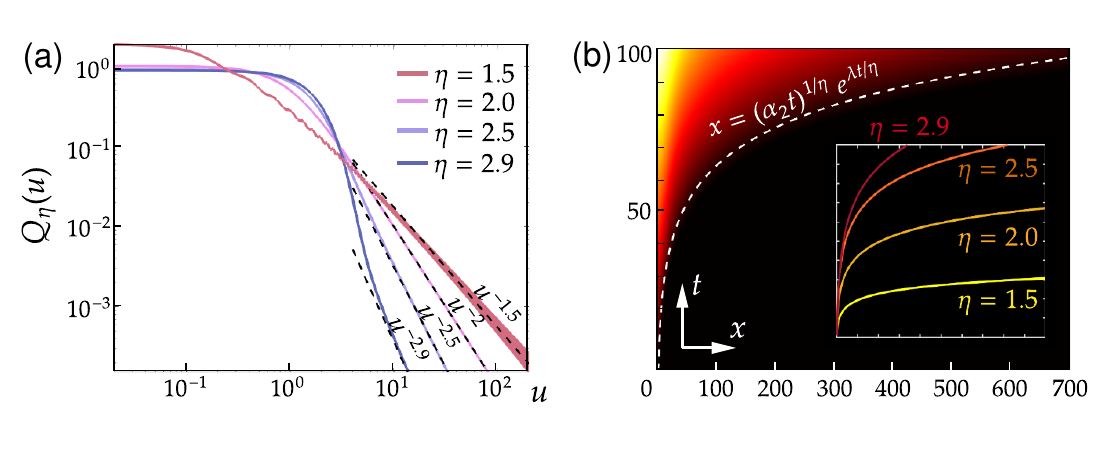}
    \caption{(a) Verification of the integral $Q_{\eta}(u)$, with dashed lines representing $u^{-\eta}$ and solid lines representing the numerical results. (b) Numerical calculations of the correlator and the analytical solution of the light-cone. The lower-right subfigure shows light-cones for various values of $\eta$. We set $J=0.2$ globally and $\eta=2.5$ for the correlator calculation. }
    \label{fig:int-veri}
\end{figure}
For $\eta> 3$, the late-time leading term is $\kappa^2$, which is a simple Gaussian integral giving rise to a linear light-cone. For $\eta<3$, the $|k|^{\eta-1}$ term dominates, leading to the solution
\begin{equation}
    \frac{C(x,t)}{C(0,0)}=\frac{e^{\lambda t}}{\pi(\alpha_2t)^{\frac{1}{\eta-1}}}\int_0^{\infty}\!d\kappa \,e^{-\kappa^{\eta-1}}\!\cos{\kappa u}
    \equiv
    \frac{e^{\lambda t}}{\pi(\alpha_2t)^{\frac{1}{\eta-1}}}Q_{\eta}(u),
\end{equation}
where $Q_{\eta}(u)$ is the Fourier transform of the stretched exponential function $\exp(-|k|^{\eta-1})$, which has been studied and summarized in~\cite{montroll1984levy,wuttke2012laplace}. For the special case where $\eta-1=2$, the integral corresponds to the Fourier transform of the Gaussian distribution. For general fractional $\eta<3$, the small $u$ and large $u$ expansions of the integral are
\begin{equation}\label{eq:smallu}
    Q_\eta(u)=\frac{1}{\eta-1}\sum_{n=0}^{\infty}\frac{(-1)^n}{(2n+1)!}\Gamma\qty(\frac{2n+1}{\eta-1})u^{2n},\quad u\ll1;
\end{equation}
\begin{equation}\label{eq:largeu}
    Q_\eta(u)=\sum_{m=1}^{\infty}\frac{(-1)^{m-1}}{(m+1)!}\sin\qty(\frac{m\pi(\eta-1)}{2})\Gamma(m\eta-m+1)u^{-m(\eta-1)-1},\quad u\gg 1.
\end{equation}
For sufficiently large $u$, only the first term of (\ref{eq:largeu}) matters, leading to a $Q_{\eta}(u)\sim u^{-\eta}$ tail of the integral. Since the sine term vanishes for odd $\eta$, there is no such tail in the $\eta=3$ case. Numerical integrals confirm the analytical result above, as shown in Fig. \ref{fig:int-veri}(a).

Once we have determined the long-wavelength limit of the correlator, the light-cone can be obtained by forcing $C(x,t)/C(0,0)=o(1)$, or setting $C(x,t)/C(0,0)=h_{\eta}/\pi$, where $h_{\eta}\equiv-\cos(\pi\eta/2)\Gamma(\eta)/2$ as an example. This gives us
\begin{equation}
    \frac{C(x,t)}{C(0,0)}=\frac{h_{\eta}e^{\lambda t}}{\pi(\alpha_2t)^{\frac{1}{\eta-1}}}\frac{(\alpha_2t)^{\frac{\eta}{\eta-1}}}{x^{\eta}}=\frac{h_{\eta}}{\pi}\frac{\alpha_2 t}{x^{\eta}}e^{\lambda t}=\frac{h_{\eta}}{\pi} \Longrightarrow x=\qty(\alpha_2t)^{\frac{1}{\eta}}e^{\frac{\lambda t}{\eta}}.
\end{equation}
Taking the logarithm of both sides, we obtain
\begin{equation}
    \lambda t+\log \alpha_2t-\eta\log x=0.
\end{equation}
Since the light-cone satisfies the logarithmic structure $t\sim \log x$ in the late-time limit, the large $u$ assumption is self-consistent. However, according to the expansion (\ref{eq:smallu}), $Q_{\eta}(u)$ for small $u$ (or $u\sim o(1)$) is finite, which leads to an exponentially increasing correlator.

\end{document}